\documentclass[twocolumn,secnumarabic,preprintnumbers,nofootinbib,amsmath,amssymb]{revtex4}
\usepackage[english]{babel}
\usepackage{epsfig,amsfonts}
\usepackage{bm}

\begin{document}

\title{Calculating the  Hawking Temperatures of Conventional Black Holes in the f(R) Gravity Models with the RVB Method}%

\author{Wen-Xiang Chen}
\affiliation{Department of Astronomy, School of Physics and Materials Science, Guangzhou University, Guangzhou 510006, China}
\author{Jun-Xian Li}
\affiliation{Department of Astronomy, School of Physics and Materials Science, Guangzhou University, Guangzhou 510006, China}
\author{Jing-Yi Zhang}
\email{zhangjy@gzhu.edu.cn}
\affiliation{Department of Astronomy, School of Physics and Materials Science, Guangzhou University, Guangzhou 510006, China}

\begin{abstract}
This paper attempted to apply the RVB method for calculating the Hawking temperatures of black holes under f(R) gravity. In calculating the Hawking temperature, we found a difference in the integration constant between the RVB and general methods. 

\end{abstract}

\maketitle

\section{{Introduction}}
In the classical view, a black hole is always considered an extreme object from which nothing can escape. Hawking and Bekenstein discovered that black holes can have temperature and entropy and that a black hole system can be considered a thermodynamic system. The topological properties of black holes can be characterized by topologically invariant Euler characteristic\cite{1,2,3,4,5,6,7,8,9}. Black holes have some important features that can be studied more easily by calculating the Euler characteristic. For example, the black hole entropy has been discussed before\cite{9,10}. While Padmanabhan et al. emphasized the importance of the topological properties of the horizon temperature\cite{11,12}. Recently, Robson, Villari, and Biancalana  \cite{12,13,14} showed that the Hawking temperature of a black hole is closely related to its topology. Besides, a topological method related to the Euler characteristic was proposed to obtain the Hawking temperature, which has been successfully applied to four-dimensional Schwarzschild black holes, anti-de Sitter black holes, and other Schwarzschild-like or charged black holes. Based on the work of Robson, Villari, and Biancalana \cite{12,13,14}, Liu et al. accurately calculated the Hawking temperature of a charged rotating BTZ black hole by using this topology method\cite{15}. Xian et al. \cite{16} studied the Hawking temperature of the global monopole spacetime which was based on the topological method proposed by the RVB method. Previous work has linked the Hawking temperature to the topological properties of black holes.

The previous literature applied the RVB method to various black holes under general relativity\cite{12,13,14,15,16}. Therefore, whether the RVB complex coordinate system makes their temperature difficult to calculate for many special black holes, it is found that the Hawking temperature of the black hole under f(R) gravity can be easily obtained by the RVB method. In this paper, by comparing the RVB method and the regular one, the Hawking temperatures of four types of black holes under different f(R) gravity are investigated. We found an integration constant for the temperature calculation after using the RVB method. The constant is zero or a parameter term that would otherwise not correspond to the Hawking temperature under the conventional method.

The arrangement of this paper is as follows. In the second part, the main formula studied in this paper is introduced, which are, new expressions for the Hawking temperature of a two-dimensional black hole system. In the third part, by studying the properties of known topological invariants and the RVB method, the Hawking temperature of Schwarzschild-like black holes under f(R) gravity is studied. In the fourth part, the RVB method is continued to calculate the Hawking temperature in RN black hole systems under f(R) gravity. In the fifth part, this formula is used to study the characteristics of Hawking temperature in terms of BZT black holes under f(R) gravity. In the sixth part, the RVB method is used to calculate the Hawking temperature of the Kerr-Sen black hole. The seventh part is the conclusion and discussion.

\section{General Calculation Method and RVB Method for the Hawking Temperature of a Black Hole under the f(R) Gravitational Model}
\subsection{General Method for Calculating the Temperature of a Black Hole}
In the f(R) theory, the metric of a general static spherically symmetric black hole\cite{17,18,19,20,21,22,23,24,25,26}is 
\begin{equation}
d s^{2}=-g(r) d t^{2}+\frac{d r^{2}}{n(r)}+r^{2} d \Omega^{2},
\end{equation}
where $g(r)$ and  $n(r)$ are general functions of the coordinate r. Taking $n\left(r_{+}\right)=0$ , the resulting horizon satisfies $n^{\prime}\left(r_{+}\right) \neq 0$.
\begin{equation}
\kappa_{K}=\frac{\sqrt{g^{\prime}\left(r_{+}\right) n^{\prime}\left(r_{+}\right)}}{2}.
\end{equation}
Then the black hole temperature is
\begin{equation}
T=\frac{\kappa_{K}}{2 \pi}=\frac{\sqrt{g^{\prime}\left(r_{+}\right) n^{\prime}\left(r_{+}\right)}}{4 \pi}.
\end{equation}

\subsection{RVB Method for Calculating the Temperature of a Black Hole}
Black hole systems are various, and stationary or rotating black holes have simple metrics, thus the temperature can be easily deduced. However, the complex coordinate system makes their temperature difficult to calculate for many special black holes. Therefore, the RVB method coordinates the Hawking temperature of a black hole to the Euler characteristic $\chi$, which is very useful in calculating the Hawking temperature in any coordinate system. Note: Since we are between the cosmological and event horizons, the inner horizon temperature cannot be observed. Below we discuss only the observable Hawking temperature.

The Euler characteristic\cite{1,2,3,4,5} can be described: 
\begin{equation}
\chi=\int_{r_{0}} \Pi-\int_{r_{\mathrm{H}}} \Pi-\int_{r_{0}} \Pi=-\int_{r_{\mathrm{H }}} \Pi.
\end{equation}
In a word, in the calculation of the Euler characteristic, the outer boundary is always canceled out. Therefore, the integral should only be related to the Killing horizon.

According to references \cite{12,13,14,15,16,19}, the Hawking temperature of a two-dimensional black hole can be obtained by the topological formula:
\begin{equation}
T_{H}=\frac{\hbar c}{4 \pi \chi k_{B}} \sum_{j \leq \chi} \int_{r_{H_{j}}} \sqrt{|g| }R d r,
\end{equation}
among them, $\hbar$ is Planck's constant, $c$ is the speed of light, $k_{B}$  is the Boltzmann constant, g is the determinant of the metric, R is the Ricci scalar, and $r_{H_{j} }$ is the location of the  Killing horizon. In this paper, we use the natural unit system $\hbar=c=k_{B}=1$. The $\chi$, which depends on the spatial coordinate r, is an Euler characteristic, representing the Killing level in Euclidean geometry. The symbol Eq.(4)  represents the summation associated with the horizon. Through transformation, in this paper $|g|$=1.

The complex coordinate system makes its temperature difficult to calculate many special black holes are difficult to calculate for many special black holes' theorem, it is also equal to the integral of $\Pi$ over the boundary of $V^{n}$. Bounded manifolds require an important correction to the value of $\chi$, which becomes:\cite{12,13,14}
\begin{equation}
\chi=\int_{\partial V} \Pi-\int_{\partial M} \Pi.
\end{equation}
The submanifold $V^{n}$ of $M^{2 n-1}$ is crucial because its boundaries are defined as fixed points (zero point) of the unit vector field defined in  $M^{n}$. From$|g|$=1, the Hawking temperature can be rewritten by using Eq.(5) and Eq.(6) as
\begin{equation}
T_{\mathrm{H}}=-\frac{1}{2}(\frac{1}{4 \pi} \int_{r_{c(-)}} R d r-\frac{1}{4 \pi} \int_{r_{+}} R d r),
\end{equation}
where $r_{c(-)}$ is the radius of the additional horizon(cosmological horizon or inner horizon), $r_{+}$ is the radius of the event horizon.In the Euclidean coordinate, the period $\tau$ is already fixed on the event horizon, and therefore the cosmological horizon has a conic singularity and one has to remove it by introducing a boundary.

The Einstein-Hilbert action $\sqrt{-g} R$ is a total derivative and $\sqrt{-g}$ here is 1 . Therefore the Ricci scalar of the metric (1) must be a total derivative, and the only thing that I can write must be $R \sim -g^{\prime \prime}$. To be precise, it is
\begin{equation}
R\sim -g^{\prime \prime}(r) .
\end{equation}
Therefore
\begin{equation}
\int d r R=-g^{\prime}(r),\left.\quad \rightarrow \quad\left(\int d r R\right)\right|_{r \rightarrow r_{+}} \sim -g^{\prime}\left(r_{+}\right)+C.
\end{equation}
This leads to the relation to the Hawking temperature. The more general case works as well, with
\begin{equation}
d s^2=-g(r) d t^2+\frac{d r^2}{n(r)}, \quad \rightarrow \quad \sqrt{-g} R \sim-\left(\sqrt{\frac{n}{g}} g^{\prime}\right)^{\prime}.
\end{equation}

\section{Hawking Temperature of Schwarzschild-like black holes under f(R) gravity obtained by RVB method}
The f(R) static black hole solution and its thermodynamics (when the constant Ricci curvature is not equal to 0) are briefly reviewed. Its general form of action is given by
\begin{equation}
I=\frac{1}{2} \int d^{4} x \sqrt{-g} f(R)+S_{\text {mat }}.
\end{equation}

\subsection{In the case of constant Ricci curvature (initial conditions)}
 
\subsubsection{Schwarzschild-de Sitter-f(R) black holes}
A spherically symmetric solution with constant curvature $R_{0}$ is first considered as a simple but important example. By comparison with \cite{27,28,29,30,31,32,33,34,35,36}, the Schwarzschild solution ($R_0$ = 0)or the Schwarzschild-de Sitter solution is
\begin{equation}
d s^{2}=-g(r) d t^{2}+\frac{d r^{2}}{g(r)}+r^{2} d \Omega^{2},
\end{equation}
where
\begin{equation}
g(r)=1-\frac{2 M}{r}-\frac{R_{0} r^{2}}{12}.
\end{equation}
It is not difficult to calculate a Schwarzschild spherical symmetric solution with constant curvature $R_{0}$ under the f(R) model, the surface gravity on the event horizon and the cosmological horizon are
\begin{equation}
\begin{aligned}
&\kappa_{\mathrm{h}}=\frac{R_{0}}{48} r_{\mathrm{+}}^{-1}\left(r_{\mathrm{c}}-\mathrm{r}_{\mathrm{+}}\right)\left(\mathrm{r}_{\mathrm{+}}-\mathrm{r}_{-}\right), \\
&\kappa_{\mathrm{c}}=\frac{R_{0}}{48} \mathrm{r}_{\mathrm{c}}^{-1}\left(\mathrm{r}_{\mathrm{c}}-\mathrm{r}_{\mathrm{+}}\right)\left(\mathrm{r}_{\mathrm{c}}-\mathrm{r}_{-}\right) ,
\end{aligned}
\end{equation}
where $R_{0}$ is the constant curvature,$r_{c}$ is the radius of the cosmological horizon, $r_{+}$ is the radius of the event horizon, and $r_{-}$ is the radius of the inner horizon.

It can be seen that the Schwarzschild-de Sitter-f(R) black hole has two horizons, both of which have Hawking radiation.
In the Cartesian coordinate system, the two-dimensional line element is\cite{10,17}
\begin{equation}
d s^{2}=g(r) d \tau^{2}+\frac{d r^{2}}{g(r)}.
\end{equation}
Thus the Ricci scalar is
\begin{equation}
R=\frac{d^{2}}{d r^{2}} g(r)=-\frac{4 M}{r^{3}}-1/6R_{0}.
\end{equation}
From $|g|$=1, the Hawking temperature is gotten by Eq.(5) and Eq.(6):
\begin{equation}
T_{\mathrm{H}}=-\frac{1}{2}\left(\frac{1}{4 \pi} \int_{r_{c}} R d r-\frac{1}{4 \pi} \int_{r_{+}} R d r\right).
\end{equation}
There were two Killing horizons at that time, the Euler characteristic is 2, and we get
\begin{equation}
T_{H}=\kappa_{\mathrm{h}}/(2\pi)+\kappa_{\mathrm{c}}/(2\pi)+C.
\end{equation}
$r_{c}$ is the radius of the cosmological horizon, $r_{+}$ is the radius of the event horizon, both are the radius of the Killing horizon, and C is the integral constant. In the Euclidean coordinate, the period $\tau$ is already fixed on the event horizon, and therefore the cosmological horizon has a conic singularity and one has to remove it by introducing a boundary.

Comparing Eq.(18) with the Hawking temperature under the conventional method
\begin{equation}
T_{H}=\kappa_{\mathrm{h}}/(2\pi)+\kappa_{\mathrm{c}}/(2\pi),
\end{equation}
we noticed C is 0.

\subsubsection{Black hole in the form of f(R) theory: $f(R)=R-q R^{\beta+1} \frac{\alpha \beta+\alpha+\epsilon}{\beta+1}+q \epsilon R ^{\beta+1} \ln \left(\frac{a_{0}^{\beta} R^{\beta}}{c}\right)$}
One of the forms of f(R) gravity is\cite{17,18,19,20,21,22,23,24,25,26}
\begin{equation}
f(R)=R-q R^{\beta+1} \frac{\alpha \beta+\alpha+\epsilon}{\beta+1}+q \epsilon R^{\beta+1} \ln \left(\frac{a_{0}^{\beta} R^{\beta}}{c}\right),
\end{equation}
where $0 \leq \epsilon \leq \frac{e}{4}\left(1+\frac{4}{e} \alpha\right), q=4 a_{0}^{\beta} / c( \beta+1), \alpha \geq 0, \beta \geq 0$ and $a_{0}=l_{p}^{2}, a$ and $c$ are constants. Since $R \neq 0$, this $f(R)$ theory has no Schwarzschild solution. Its metric form is
\begin{equation}
d s^{2}=-g(r) d t^{2}+h(r) d r^{2}+r^{2} d \theta^{2}+r^{2} \sin ^{2 } \theta d \varphi^{2},
\end{equation}
and
\begin{equation}
g(r)=h(r)^{-1}=1-\frac{2 m}{r}+\beta_{1} r ,
\end{equation}
where $\mathrm{m}$ is related to the mass of the black hole, and $\beta_{1}$ is a model parameter.

Since the definition of the Killing horizon does not involve angular degrees of freedom, we reduce the angular degrees of freedom of space-time. In the Cartesian coordinate system, the two-dimensional metric is as follow\cite{10,17}, and the Euler characteristic is 1:
\begin{equation}
d s^{2}=g(r) d \tau^{2}+\frac{d r^{2}}{g(r)}.
\end{equation}
Thus the Ricci scalar is
\begin{equation}
R=\frac{d^{2}}{d r^{2}} g(r)=-\frac{4 m}{r^{3}}.
\end{equation}
From $|g|$=1, the Hawking temperature writes
\begin{equation}
T_{H}=\frac{2 m}{{r}_{+}^{2}}/(4\pi)+C.
\end{equation}C is the integral constant, ${r}_{+}$ is the radius of the event horizon, which is also the only radius of the Killing horizon,
\begin{equation}
r_+=\frac{\sqrt{8 \beta_{1} m+1}-1}{2 \beta_{1}}.
\end{equation}
The Hawking temperature under the conventional method is
\begin{equation}
T_{H}=(\frac{2 m}{{r}_{+}^{2}}+\beta_{1})/(4\pi),
\end{equation}we noticed C is $\beta_{1}/(4\pi)$.

\subsection{In the case of non-constant Ricci curvature (initial conditions)}

\subsubsection{Black hole in the form of f(R) theory: $f(R)=R+2 \alpha \sqrt{R}$}
Its metric is\cite{30} 
\begin{equation}
d s^{2}=-g(r) d t^{2}+1/g(r) d r^{2}+r^{2} d \theta^{2}+r^{2} \sin ^{2 } \theta d \varphi^{2},
\end{equation}
where
\begin{equation}
g(r)=\frac{1}{2}+\frac{1}{3 \alpha r}.
\end{equation}

The Hawking temperature is gotten by the use of the RVB method. Euler characteristic is 1, and the Killing horizon is the event horizon. In the Cartesian coordinate system, the two-dimensional line element is 
\cite{10,17}
\begin{equation}
d s^{2}=g(r) d \tau^{2}+\frac{d r^{2}}{g(r)}.
\end{equation}
Thus the Ricci scalar is
\begin{equation}
R=2 /(3 \alpha (r^3)).
\end{equation}
The Hawking temperature writes
\begin{equation}
T_{H}=-1/((12\pi)\alpha)r_{+}^2)+C.
\end{equation}C is the integral constant.At the event horizon, $\mathrm{g}\left(\mathrm{r}_{+}\right)=0$, resulting in
\begin{equation}
 r_+=-\frac{2}{3\alpha}.
\end{equation}
The Hawking temperature under the conventional method is
\begin{equation}
T_{H}=-1/((12\pi)\alpha)r_{+}^2).
\end{equation}In contrast, C is 0.

\subsubsection{Black hole in the form of f(R) theory: $f(R)=R+2 \alpha \sqrt{R-4 \Lambda}-2 \Lambda$}
The second f(R) model considered with non-constant curvature is given by literature \cite{17,18,19,20,21,22,23,24,25,26,27,28}
\begin{equation}
f(R)=R+2 \alpha \sqrt{R-4 \Lambda}-2 \Lambda,
\end{equation}
where $\alpha<0$  is the integral constant, which can be regarded as the cosmological constant. The solution to this model is as follows
\begin{equation}
d s^{2}=-g(r) d t^{2}+1/g(r) d r^{2}+r^{2} d \theta^{2}+r^{2} \sin ^{2 } \theta d \varphi^{2}.
\end{equation}

 In the Cartesian coordinate system, the two-dimensional line element is\cite{10,17}
\begin{equation}
d s^{2}=g(r) d \tau^{2}+\frac{d r^{2}}{g(r)},
\end{equation}
where
\begin{equation}
g(r)=\frac{1}{2}+\frac{1}{3 \alpha r}-\frac{\Lambda}{3} r^{2}.
\end{equation}

There is only one positive root of g(r) = 0, so the Ricci scalar is
\begin{equation}
R=2 /(3 \alpha (r^3))-2 \Lambda/3.
\end{equation}
Using the RVB method, we see that
\begin{equation}
T_{H}=-1/((12\pi)\alpha)r_{+}^2)-2 \Lambda r_{+}/(12\pi)+C.
\end{equation}C is the integral constant.

The Hawking temperature under the conventional method is
\begin{equation}
T_{H}=-1/((12\pi)\alpha)r_{+}^2)-2 \Lambda r_{+}/(12\pi).
\end{equation}In contrast, C is 0.

In another situation, when the cosmological constant is positive, the Euler characteristic is 2, the metric is isomorphic to the example in 3.1.1, and the conclusion is consistent with 3.1.1.

\subsubsection{Black hole in the form of f(R) theory: $f(R)=R-\frac{\mu^{4}}{R}$ and $f(R)=R-\lambda \exp (-\xi R)$}
The solution to this model is:\cite{30,31,32,33,34,35,36}
\begin{equation}
d s^{2}=-g(r) d t^{2}+1/g(r) d r^{2}+r^{2} d \theta^{2}+r^{2} \sin ^{2 } \theta d \varphi^{2}.
\end{equation}
 And when the cosmological constant is negative,d=4,
\begin{equation}
g(r)=k_c-\frac{2 \Lambda}{(d-1)(d-2)} r^{2}-\frac{M}{r^{d-3}},
\end{equation}where we should set $\Lambda=\frac{\pm \mu^{2}}{2 d}\sqrt{d^{2}-4}$ or $\lambda=\frac{2 d \Lambda e^{\xi R}}{d+2 \xi R}$.${k_c}$ is a constant, it can take 1, -1, 0.
We should set $\Lambda$ as the negative terminal (d as the dimension). In the Cartesian coordinate system, the two-dimensional line element is\cite{10,17}
\begin{equation}
d s^{2}=g(r) d \tau^{2}+\frac{d r^{2}}{g(r)}.
\end{equation}
Thus the Ricci scalar is
\begin{equation}
R=\frac{d^{2}}{d r^{2}} g(r)=-(d-3)(d-2)\frac{M}{r^{d-1}}-4/({(d-1)(d-2)})\Lambda.
\end{equation}
The Hawking temperature writes
\begin{equation}\resizebox{1\hsize}{!}{$
T_{H}=-1/((d-1)(d-2)\pi)\Lambda r_{+}+(d-3)M/ (4\pi (r_{+})^{d-2})+C.$}
\end{equation}C is the integral constant.

Let g(r)=0, we get
\begin{equation}\resizebox{1\hsize}{!}{$
\begin{aligned}
    r_+ & =-((2^{1 / 3} \times(2 k-3 d k+d^{2} k)) /(216 M \Lambda^{2}-324 d M \Lambda^{2}+108 d^{2} M \Lambda^{2}+\\
    &\sqrt{-864(2 k-3 d k+d^{2} k)^{3} \Lambda^{3}+(216 M \Lambda^{2}-324 d M \Lambda^{2}+108 d^{2} M \Lambda^{2})^{2}})^{1 / 3})\\
    &-\frac{1}{6 \times 2^{1 / 3} \Lambda}(216 M \Lambda^{2}-324 d M \Lambda^{2}+108 d^{2} M \Lambda^{2}\\
    &+\sqrt{-864(2 k-3 d k+d^{2} k)^{3} \Lambda^{3}+\left(216 M \Lambda^{2}-324 d M \Lambda^{2}+108 d^{2} M \Lambda^{2}\right)^{2}})^{1 / 3}.
    \end{aligned}$}
\end{equation}

The Hawking temperature under the conventional method is
\begin{equation}
T_{H}=-1/((d-1)(d-2)\pi)\Lambda r_{+}+(d-3)M/ (4\pi (r_{+})^{d-2}).
\end{equation}By comparison, we get C is 0.

In another situation-when the cosmological constant is positive, the Euler characteristic is 2, the metric is isomorphic to the example in 3.1.1, and the conclusion is consistent with 3.1.1.

\subsubsection{Black hole in the form of f(R) theory: $df(R)/dR=1+\alpha r$ }
The line element is\cite{36}
\begin{equation}
d s^{2}=-g(r) d t^{2}+1/g(r) d r^{2}+r^{2} d \theta^{2}+r^{2} \sin ^{2 } \theta d \varphi^{2},
\end{equation}where
\begin{equation}
\resizebox{1\hsize}{!}{$
g(r)=C_{2} r^{2}+\frac{1}{2}+\frac{1}{3 \alpha r}+\frac{C_{1}}{r}\left[3 \alpha r-2-6 \alpha^{2} r^{2}+6 \alpha^{3} r^{3} \ln \left(1+\frac{1}{\alpha r}\right)\right],$}
\end{equation}
C1 and C2 are constants.

Using the RVB method, we see that the Killing horizon is the event horizon. In the Cartesian coordinate system, the two-dimensional metric is\cite{10,17}
\begin{equation}
d s^{2}=g(r) d \tau^{2}+\frac{d r^{2}}{g(r)}.
\end{equation}
So the Ricci scalar is
\begin{equation}
\begin{aligned}
    R&=2C_2+\frac{2}{3ar^{3}}-\frac{4C_1}{r^{3}}-C_1\frac{12 a^{2}}{\left(\frac{1}{a r}+1\right) r}\\
    &-\frac{6 aC_1}{\left(\frac{1}{a r}+1\right)^{2} r^{2}}+12 a^{3}C_1 \ln \left(\frac{1}{a r}+1\right).
\end{aligned}
\end{equation}
The Hawking temperature is
\begin{equation}\resizebox{0.9\hsize}{!}{$
\begin{aligned}
    T_{H} & =(2C_2  r_{+}- \frac{1}{3 a r_{+}^{2}}+ \frac{2 C_1}{ r_{+}^{2}}+C_1(12 a^{3} \ln \left(\frac{1}{a r_{+}}+1\right) r_{+}\\
    &-\frac{6 a^{2}}{\frac{1}{a r_{+}}+1}))/ (4\pi)+C.
\end{aligned}$}
\end{equation}
C is the integral constant. The Hawking temperature under the conventional method is
\begin{equation}\resizebox{0.9\hsize}{!}{$
\begin{aligned}
    T_{H} & =(2C_2  r_{+}- \frac{1}{3 a r_{+}^{2}}+ \frac{2 C_1}{ r_{+}^{2}}+C_1(12 a^{3} \ln \left(\frac{1}{a r_{+}}+1\right) r_{+}\\
    &-\frac{6 a^{2}}{\frac{1}{a r_{+}}+1}))/ (4\pi).
\end{aligned}$}
\end{equation}
In contrast, C is $-6 a^{2}C_1/ (4\pi)$.

In another situation, when g(r) = 0, it can have two more positive roots, the metric is isomorphic to the example in 3.1.1, and the conclusion is consistent with 3.1.1.

\subsubsection{Black hole in the form of f(R) theory: $f(R)=R+\Lambda+\frac{R+\Lambda}{R / R_{0}+2 / \alpha} \ln \frac{R+\Lambda}{R_{c}}$}
The line element is\cite{37}
\begin{equation}
d s^{2}=-g(r) d t^{2}+1/g(r) d r^{2}+r^{2} d \theta^{2}+r^{2} \sin ^{2 } \theta d \varphi^{2},
\end{equation}where
\begin{equation}
g(r)=1-\frac{2 M}{r}+\beta r-\frac{\Lambda r^{2}}{3}.
\end{equation}
$\beta>0$.

We use the RVB method to find the Hawking temperature. At this time, the Euler characteristic is 1, and the event horizon is the only Killing horizon.
In the Cartesian coordinate system, the two-dimensional  metric is \cite{10,17}
\begin{equation}
d s^{2}=g(r) d \tau^{2}+\frac{d r^{2}}{g(r)}.
\end{equation}
So the Ricci scalar is
\begin{equation}
R=-\frac{4 M}{r^{3}}-\frac{2\Lambda}{3}.
\end{equation}When g(r) = 0 has only one positive root, the Euler characteristic is 1,
\begin{equation}
T_{H}=(\frac{2 M}{r_{+}^{2}}-\frac{2 \Lambda r_{+}}{3})/ (4\pi)+C.
\end{equation}
C is the integral constant.

Let g(r)=0, we get
\begin{equation}\resizebox{0.9\hsize}{!}{$
\begin{aligned}
    r_+ &= \frac{\beta}{\Lambda}+\frac{2^{1 / 3} \times\left(-9 \beta^{2}-9 \Lambda\right)}{3 \Lambda\left(-54 \beta^{3}-81 \beta \Lambda+162 M \Lambda^{2}+\sqrt{4\left(-9 \beta^{2}-9 \Lambda\right)^{3}+\left(-54 \beta^{3}-81 \beta \Lambda+162 M \Lambda^{2}\right)^{2}}\right)^{1 / 3}}-\\
    &\frac{\left(-54 \beta^{3}-81 \beta \Lambda+162 M \Lambda^{2}+\sqrt{4\left(-9 \beta^{2}-9 \Lambda\right)^{3}+\left(-54 \beta^{3}-81 \beta \Lambda+162 M \Lambda^{2}\right)^{2}}\right)^{1 / 3}}{3 \times 2^{1 / 3} \Lambda}.
\end{aligned}$}
\end{equation}

The Hawking temperature under the conventional method is
\begin{equation}
T_{H}=(\frac{2 M}{r_{+}^{2}}-\frac{2 \Lambda r_{+}}{3}+\beta)/ (4\pi).
\end{equation}
By contrast, C is $\beta/ (4\pi)$.

When the cosmological constant is positive, the metric is isomorphic to the example in 3.1.1, the conclusion is consistent with 3.1.1, and the Euler characteristic is 2.

\section{Hawking temperature of RN black holes under f(R) gravity by RVB method}
In this section, we briefly review the main features of four-dimensional charged black holes with constant Ricci scalar curvature \cite{13,31,37,38,39} in the f(R) gravitational background.  The action  is given by
\begin{equation}
S=\int_{\mathcal{M}} d^{4} x \sqrt{-g}\left[f(R)-F_{\mu \nu} F^{\mu \nu}\right].
\end{equation}

\subsection{Solutions of the black hole with constant Ricci curvature}
\subsubsection{$f(R)-RN-de Sitter$ black hole}
Comparing the Reissner Nordstrom black hole in the de Sitter space-time, we consider  the spherically symmetric solution of $f(R)$ gravity with constant curvature $R_{0}$(or initial Ricci curvature) under the model, such as RN black hole solution $\left(R_{0}=0\right)$ , where $g(r)$ is in the form of $g(r )=1-\frac{2 M}{r}-\frac{R_{0} r^{2}}{12}+\frac{Q^2}{r^2}$, its  metric is as follows \cite{33}
\begin{equation}
\begin{aligned}
    \mathrm{d}s^{2} &=-\left(1-\frac{2 M}{r}-\frac{R_{0} r^{2}}{12}+\frac{Q^2 }{r^2}\right) \mathrm{d} t^{2} \\
    &+\left(1-\frac{2 M}{r}-\frac{R_{0} r^{2}}{12}+\frac{Q^2}{r^2}\right)^{- 1} \mathrm{~d} r^{2} \\
    &+r^{2} d \theta^2+r^{2} \sin^{2}{\theta}d\varphi^2,
\end{aligned}
\end{equation}
where $R_0(>0)$ is the cosmological constant.

Let g(r)=0, three solutions can be obtained,
\begin{equation}
1-\frac{2 M}{r}-\frac{R_{0} r^{2}}{12}+\frac{Q^2}{r^2}=0.
\end{equation}
 $r_1$ is a negative root, which has no physical meaning; $r_{i}$ is the smallest positive root, corresponding to the inner horizon of the black hole; $r_{\mathrm{e}}$ is a smaller positive root, corresponding to the outer horizon of the black hole; $r_{\mathrm{c}}$ is the largest positive root. The three surface gravities are 
\begin{equation}
\begin{gathered}
\kappa1=\frac{R_{0}}{48} r_{i}^{-2} (\begin{array}{lll}r_{i} -r_1\end{array} ) (\begin{array} {lll}r_{\mathrm{e}} -r_{\mathrm{i}}\end{array} ) (\begin{array}{lll}r_{\mathrm{c}} -r_{\mathrm{i }}\end{array} ),\\
\kappa2=\frac{R_{0}}{48} r_{\mathrm{e}}^{-2}\left(r_{\mathrm{e}}-\mathrm{r_1}\right)\left( r_{\mathrm{e}}-r_{\mathrm{i}}\right)\left(r_{\mathrm{c}}-r_{\mathrm{e}}\right),\\
\kappa3=\frac{R_{0}}{48} r_{\mathrm{c}}^{-2}\left(r_{\mathrm{c}}-\mathrm{r_1}) \left(r_{ \mathrm{c}}-r_{\mathrm{i}}\right)\left(r_{\mathrm{c}}-r_{\mathrm{e}}\right) .\right.
\end{gathered}
\end{equation}
 We use the RVB method to find the Hawking temperature.
At this time, the Euler characteristic is 2, and we get, in the Cartesian coordinate system, the two-dimensional metric is \cite{10,17}
\begin{equation}
d s^{2}=g(r) d \tau^{2}+\frac{d r^{2}}{g(r)},
\end{equation}
so the Ricci scalar is
\begin{equation}
R=\frac{d^{2}}{d r^{2}} g(r)=-\frac{4 M}{r^{3}}-1/6R_{0}+6Q^2/r ^4.
\end{equation}
Using Eq.(7), we get
\begin{equation}
T_{\mathrm{H}}=-\frac{1}{2}\left(\frac{1}{4 \pi} \int_{r_{c}} R d r-\frac{1}{4 \pi} \int_{r_{e}} R d r\right),
\end{equation}
and
\begin{equation}
T_{H}=\kappa_{\mathrm{2}}/(2\pi)+\kappa_{\mathrm{3}}/(2\pi)+C.
\end{equation}$r_{c}$ is the radius of the cosmological horizon, $r_{e}$ is the radius of the event horizon, both of which are the radius of the Killing horizon, and C is an integral constant. Since we are between the cosmological and event horizons, the inner horizon temperature cannot be observed. Below we will only talk about the observable Hawking temperature. In the Euclidean signature, the period $\tau$ is already fixed on the event horizon, and therefore the cosmological horizon has a conic singularity and one has to remove it by introducing a boundary.

The Hawking temperature under the conventional method is
\begin{equation}
T_{H}=\kappa_{\mathrm{2}}/(2\pi)+\kappa_{\mathrm{3}}/(2\pi).
\end{equation}By contrast, C is 0.

\subsubsection{Black hole in the form of f(R) theory: $f(R)=R-\alpha R^{n}$}
The example we focus on is \cite{16,30}
\begin{equation}
f(R)=R-\alpha R^{n}.
\end{equation}
Its metric form is
\begin{equation}
d s^{2}=-g(r) d t^{2}+\frac{d r^{2}}{g(r)}+r^{2} d \Omega^{2},
\end{equation}
where
\begin{equation}
g(r)=1-\frac{2 m}{r}+\frac{q^{2}}{b r^{2}}.
\end{equation}
$b=f^{\prime}\left(R_{0}\right)$ and the two parameters $m$ and $q$ are proportional to the black hole mass and charge, respectively \cite{14}
\begin{equation}
M=m b, \quad Q=\frac{q}{\sqrt{b}}.
\end{equation}

 In the Cartesian coordinate system, the two-dimensional line element is \cite{10,17}
\begin{equation}
d s^{2}=g(r) d \tau^{2}+\frac{d r^{2}}{g(r)},
\end{equation}
so the Ricci scalar is
\begin{equation}
R=\frac{d^{2}}{d r^{2}} g(r)=-\frac{4 m}{r^3}+\frac{6q^{2}}{b r^{4}}.
\end{equation}
We got
\begin{equation}
\begin{aligned}
T_{H}=(\frac{ m}{{r_+}^2}-\frac{q^{2}}{b {r_+}^{3}}+\frac{ m}{{r_-}^2}-\frac{q^{2}}{b {r_-}^{3}})/(4\pi) +C.
\end{aligned}
\end{equation}$r_{-}$ is the radius of the inner horizon, $r_{+}$ is the radius of the event horizon, both are the radius of the Killing horizon, and C is the integral constant.
\begin{equation}
r_+=\sqrt{m^{2}-Q^{2}}+m,r_-=m-\sqrt{m^{2}-Q^{2}}.
\end{equation}
The Hawking temperature under the conventional method is
\begin{equation}
T_{H}=(\frac{ m}{{r_+}^2}-\frac{q^{2}}{b {r_+}^{3}}+\frac{ m}{{r_-}^2}-\frac{q^{2}}{b {r_-}^{3}})/(4\pi).
\end{equation}
By contrast, C is 0.

\subsubsection{Black hole in the form of f(R) theory: $f(R)=R-\lambda \exp (-\xi R)+\kappa R^{2}$}
We define $\lambda=\frac{R e^{\xi R}}{2+\xi R}, \kappa=-\frac{1+\xi R}{R(2+\xi R)} $ , $\xi$ is a free parameter, to simplify our following $d$ takes 4.
Its metric form is\cite{35}
\begin{equation}
d s^{2}=-g(r) d t^{2}+\frac{d r^{2}}{g(r)}+r^{2} d \Omega^{2},
\end{equation}
where
\begin{equation}
g(r)=k-\frac{2 \Lambda}{(d-1)(d-2)} r^{2}-\frac{M}{r^{d-3}}+\frac{ Q^{2}}{r^{d-2}}.
\end{equation}${k}$ is a constant, it can take 1, -1, 0.
In the Cartesian coordinate system, the two-dimensional metric is \cite{10,17}
\begin{equation}
d s^{2}=g(r) d \tau^{2}+\frac{d r^{2}}{g(r)},
\end{equation}
so the Ricci scalar is
\begin{equation}\resizebox{1\hsize}{!}{$
R=\frac{d^{2}}{d r^{2}} g(r)=-(d-3)(d-2)\frac{M}{r^{d-1}}-4 /({(d-1)(d-2)})\Lambda+(d-1)(d-2)Q^2/r^d.$}
\end{equation}
Some parameters, such as $\lambda$ and $\kappa$ should be fixed. Through the metric form, we find that for this static spherically symmetric black hole, it can be easily seen that it is consistent with the calculation result of the Hawking temperature under the conventional method through the metric form (the integral constant obtained is 0). When the cosmological constant is negative, at this time the Euler characteristic number is 1, and we get 
\begin{equation}\resizebox{1\hsize}{!}{$
T_{H}=-1/((d-1)(d-2)\pi)\Lambda r_{+}+(d-3)M/ (4\pi (r_{+})^{d- 2})-(d-2)Q^2/ (4\pi (r_{+})^{d-1})+C.$}
\end{equation}C is the integral constant, and $r_{+}$ is the radius of the event horizon.

The Hawking temperature under the conventional method is
\begin{equation}\resizebox{1\hsize}{!}{$
T_{H}=-1/((d-1)(d-2)\pi)\Lambda r_{+}+(d-3)M/ (4\pi (r_{+})^{d- 2})-(d-2)Q^2/ (4\pi (r_{+})^{d-1}).$}
\end{equation}C is 0 when contrast is made.

When the cosmological constant is positive, the structure is consistent with 4.1.1, and the conclusion is the same.

\subsection{The solution of RN black hole has a non-constant Ricci curvature}
\subsubsection{Black hole in the form of f(R) theory:$f(R)=2 a \sqrt{R-\alpha}$}
One of the forms of the f(R) model is \cite{30}
\begin{equation}
f(R)=2 a \sqrt{R-\alpha},
\end{equation}
where $\alpha$ is a parameter of the model that is related to an effective cosmological constant, and  $\alpha>0$ is a parameter in units of [distance $]^{-1}$. The model has a solution of the static spherically symmetric black hole, which has the following form
\begin{equation}
d s^{2}=-g(r) d t^{2}+\frac{d r^{2}}{g(r)}+r^{2} d \Omega^{2},
\end{equation}
where
\begin{equation}
g(r)=\frac{1}{2}\left(1-\frac{\alpha r^{2}}{6}+\frac{2 Q}{r^{2}}\right),
\end{equation}
and $Q$ is the integration constant. The event horizon of the black hole is located at: $(a)$ when $\alpha>0$ and $Q>0$, $r_{+}=\sqrt{3 \alpha+\alpha \sqrt{9+12 \alpha Q}}/\alpha$;
(b) when $\alpha>0$, $Q<0$ and $\alpha Q>-3/4$, $r_{+}=\sqrt{3\alpha-\alpha \sqrt{9+12 \alpha Q}}/\alpha$; (c) when $\alpha<0$ and $Q<0$, $r_{+}=\sqrt{3/\alpha-\alpha \sqrt{9+12 \alpha Q} / \alpha}$; (d) when $\alpha>0$ and $Q=0$, $r_{+}=\sqrt{6/\alpha}$.

When  $\alpha>0$ and $Q>0$ or  $\alpha<0$ and $Q<0$ or $\alpha>0$ and $Q=0$, then g (r)=0 has only one positive root with physical meaning. The Euler characteristic is 1, and $|g|$=1 at this time.
In the Cartesian coordinate system, the two-dimensional metric is \cite{10,17}
\begin{equation}
d s^{2}=g(r) d \tau^{2}+\frac{d r^{2}}{g(r)}.
\end{equation}
The Ricci scalar is as follows
\begin{equation}
R=\frac{d^{2}}{d r^{2}} g(r)=6Q/{r^{4}}-\alpha/6.
\end{equation}
Using the RVB method, we get
\begin{equation}
T_{H}=(-\frac{2 Q}{r_{+}^{3}}-\frac{r_{+}\alpha}{6})/(4\pi)+C.
\end{equation}C is the integral constant.

The Hawking temperature under the traditional method is
\begin{equation}
T_{H}=(-\frac{2 Q}{r_{+}^{3}}-\frac{r_{+}\alpha}{6})/(4\pi).
\end{equation}In contract, C is 0.

When we take different values for $\alpha$ and $\mathrm{Q}$, we find that there are two effective positive roots in the equation $\mathrm{g}(\mathrm{r})=0$, one is the event horizon and the other is the cosmological horizon, both of which are Killing horizons. The structure is consistent with 4.1.1.  When $\alpha>0$ and $Q<0$, the Euler characteristic is 2 by this time. The RVB method is used to find the Hawking temperature:
\begin{equation}
T_{\mathrm{H}}=-\frac{1}{2}\left(\frac{1}{4 \pi} \int_{r_{c}} R d r-\frac{1}{4 \pi} \int_{r_{e}} R d r\right).
\end{equation}
\begin{equation}
T_{H}=\kappa_{\mathrm{2}}/(2\pi)+\kappa_{\mathrm{3}}/(2\pi)+C.
\end{equation}$r_{c}$ is the radius of the cosmological horizon, $r_{e}$ is the radius of the event horizon, both are the radius of the Killing horizon, and C is an integral constant. In the Euclidean signature, the period $\tau$ is already fixed on the event horizon, so the cosmological horizon has a conic singularity, which must be eliminated by introducing a boundary.

The Hawking temperature under normal conditions is
\begin{equation}
T_{H}=\kappa_{\mathrm{2}}/(2\pi)+\kappa_{\mathrm{3}}/(2\pi).
\end{equation}By contract, C is 0. This conclusion is consistent with 4.1.1,
\begin{equation}
\begin{aligned}
\kappa2=\frac{R_{0}}{48} r_{\mathrm{e}}^{-2}\left(r_{\mathrm{e}}-\mathrm{r1}\right)\left( r_{\mathrm{e}}-r_{\mathrm{i}}\right)\left(r_{\mathrm{c}}-r_{\mathrm{e}}\right),\\
\kappa3=\frac{R_{0}}{48} r_{\mathrm{c}}^{-2}\left(r_{\mathrm{c}}-\mathrm{r1}) \left(r_{ \mathrm{c}}-r_{\mathrm{i}}\right)\left(r_{\mathrm{c}}-r_{\mathrm{e}}\right) .\right.
\end{aligned}
\end{equation}
$R_{0}$ is the initial Ricci curvature.

\subsubsection{Black hole in the form of f(R) theory: $f(R)=R-\lambda \exp (-\xi R)+\kappa R^{n}+\eta \ln (R)$}
Its metric form is\cite{35}
\begin{equation}
d s^{2}=-g(r) d t^{2}+\frac{d r^{2}}{g(r)}+r^{2} d \Omega^{2},
\end{equation}
where
\begin{equation}
g(r)=k_1-\frac{2 \Lambda}{(d-1)(d-2)} r^{2}-\frac{M}{r^{d-3}}+\frac{ Q^{2}}{r^{d-2}}.
\end{equation}We take d=4,
\begin{equation}
\begin{aligned}
&\lambda=\frac{R+\kappa R^{n}-\left(R+n \kappa R^{n}\right) \ln R}{(1+\xi R \ln R) e^{ -\xi R}}, \\
&\eta=-\frac{(1+\xi R) R+(n+\xi R) \kappa R^{n}}{1+\xi R \ln R},
\end{aligned}
\end{equation}where $\xi$ is a free parameter, and ${k_1}$ is a constant, it can take 1, -1, 0.

When the cosmological constant is positive, the structure is consistent with 4.1.1, and the conclusion is the same. We find that there are two effective positive roots of the equation $g (r) =0$, one is on the event horizon and the other is on the cosmological horizon, both of which are Killing horizons.

We use the RVB method to find the Hawking temperature,
\begin{equation}
T_{\mathrm{H}}=-\frac{1}{2}\left(\frac{1}{4 \pi} \int_{r_{c}} R d r-\frac{1}{4 \pi} \int_{r_{e}} R d r\right).
\end{equation}
\begin{equation}
T_{H}=\kappa_{\mathrm{2}}/(2\pi)+\kappa_{\mathrm{3}}/(2\pi)+C.
\end{equation}In the Euclidean signature, the period $\tau$ is already fixed on the event horizon, so the cosmological horizon has a conic singularity, which must be eliminated by introducing a boundary.

The Hawking temperature under normal conditions is
\begin{equation}
T_{H}=\kappa_{\mathrm{2}}/(2\pi)+\kappa_{\mathrm{3}}/(2\pi).
\end{equation}In contrast, C is 0. This conclusion is consistent with 4.1.1,
\begin{equation}
\begin{aligned}
\kappa2=\frac{R_{0}}{48} r_{\mathrm{e}}^{-2}\left(r_{\mathrm{e}}-\mathrm{r1}\right)\left( r_{\mathrm{e}}-r_{\mathrm{i}}\right)\left(r_{\mathrm{c}}-r_{\mathrm{e}}\right),\\
\kappa3=\frac{R_{0}}{48} r_{\mathrm{c}}^{-2}\left(r_{\mathrm{c}}-\mathrm{r1}) \left(r_{ \mathrm{c}}-r_{\mathrm{i}}\right)\left(r_{\mathrm{c}}-r_{\mathrm{e}}\right) .\right.
\end{aligned}
\end{equation}$R_{0}$ is the initial Ricci curvature.

When the cosmological constant is negative, the two-dimensional metric is \cite{10,17}
\begin{equation}
d s^{2}=g(r) d \tau^{2}+\frac{d r^{2}}{g(r)},
\end{equation}
so the Ricci scalar is
\begin{equation}\resizebox{1\hsize}{!}{$
R=\frac{d^{2}}{d r^{2}} g(r)=-(d-3)(d-2)\frac{M}{r^{d-1}}-4 /({(d-1)(d-2)})\Lambda+(d-1)(d-2)Q^2/r^d.$}
\end{equation}
 When $g(r)=0$, there is only one positive root with physical meaning, at this time, the Euler characteristic is 1, and the event horizon is on the only  Killing horizon. Using the RVB method, we get
\begin{equation}\resizebox{1\hsize}{!}{$
T_{H}=-1/((d-1)(d-2)\pi)\Lambda r_{+}+(d-3)M/ (4\pi (r_{+})^{d- 2})-(d-2)Q^2/ (4\pi (r_{+})^{d-1})+C.$}
\end{equation}C is the integral constant, and $r_{+}$ is the radius of the event horizon.

 Hawking temperature under the traditional method is 
\begin{equation}\resizebox{1\hsize}{!}{$
T_{H}=-1/((d-1)(d-2)\pi)\Lambda r_{+}+(d-3)M/ (4\pi (r_{+})^{d- 2})-(d-2)Q^2/ (4\pi (r_{+})^{d-1}).$}
\end{equation}By contract, C is 0.

\section{Hawking temperature of the BTZ black hole under $f(R)$ gravity obtained by RVB method}
In the following two models, the cosmological constant is negative.
\subsection{Black hole in the form of f(R) theory: $f(R)=-4 \eta^{2} M \ln (-6 \Lambda-R)+\xi R+R_{0}$}

We bring the following solution:\cite{27,28}.
\begin{equation}
d s^{2}=-g(r) d t^{2}+\frac{d r^{2}}{g(r)}+r^{2} d \Omega^{2},
\end{equation}where
\begin{equation}
g(r)=-\Lambda r^{2}-M(2 \eta r+\xi).
\end{equation}
The curvature scalar is
\begin{equation}
R=-2 \Lambda-\frac{4 M \eta}{r}.
\end{equation}

In the Euclidean coordinate system, the two-dimensional metric is \cite{10,17}
\begin{equation}
d s^{2}=g(r) d \tau^{2}+\frac{d r^{2}}{g(r)}.
\end{equation}
Thus the Ricci scalar is
\begin{equation}
R=-2\Lambda.
\end{equation}

We use the RVB method to find the Hawking temperature,
\begin{equation}
T_{H}=\frac{-\Lambda r_{+}}{2 \pi}+C.
\end{equation}C is the integration constant.

Let g(r)=0,we get 
\begin{equation}
r_+=-\frac{\sqrt{M\left(M \eta^{2}-\Lambda \xi\right)}+M \eta}{\Lambda},
\end{equation}${r_+}$ is the event horizon radius and the unique Killing horizon radius.

The Hawking temperature calculation under the conventional method is
\begin{equation}
T=\frac{-\Lambda r_{+}-M \eta}{2 \pi}.
\end{equation} By contrast,$C= \frac{-M \eta}{2 \pi}$.

\subsection{Black hole in the form of f(R) theory: $$f(R)=-2 \eta M \ln (6 \Lambda+R)+R_{0}$$}
In the case $\Phi(r)=0$, the charged $(2+1)$-dimensional solution under pure $f(R)$-gravity is\cite{27,28}
\begin{equation}
d s^{2}=-g(r) d t^{2}+\frac{d r^{2}}{g(r)}+r^{2} d \Omega^{2},
\end{equation}
where
\begin{equation}
g(r)=-\Lambda r^{2}-M r-\frac{2 Q^{2}}{3 \eta r}.
\end{equation}
 The two-dimensional line element is\cite{10,17}
\begin{equation}
d s^{2}=g(r) d \tau^{2}+\frac{d r^{2}}{g(r)},
\end{equation} at this point $|g|$=1, the Ricci scalar is
\begin{equation}
R=-\frac{4 Q^{2}}{3 \mathrm{\eta} \mathrm{ r}^{3}}-2\Lambda.
\end{equation}
Using the RVB method, we get
\begin{equation}
T_{H}=-\frac{\Lambda r_{+}}{2 \pi}+\frac{Q^{2}}{6 \pi \eta r_{+}^{2}}+C.
\end{equation}C is the integration constant.

Let g(r)=0,we get 
\begin{equation}\resizebox{1\hsize}{!}{$
\begin{aligned}
&r_+=\frac{1}{3}\left(-\frac{M}{\Lambda}-\frac{M^{2} \eta}{\Lambda\left(M^{3} \eta^{3}+9 Q^{2} \eta^{2} \Lambda^{2}+3 \sqrt{2 M^{3} Q^{2} \ eta^{5} \Lambda^{2}+9 Q^{4} \eta^{4} \Lambda^{4}}\right)^{1 / 3}}-\right.\\\
&\left.\frac{\left(M^{3} \eta^{3}+9 Q^{2} \eta^{2} \Lambda^{2}+3 \sqrt{2 M^{3} Q^{2} \eta^{5} \Lambda^{2}+9 Q^{4} \eta^{4} \Lambda^{4}}\right) ^{1 / 3}}{\eta \Lambda}\right),
\end{aligned}$}
\end{equation}${r_+}$ is the event horizon radius and the unique Killing horizon radius.

The Hawking temperature calculation under the conventional method is 
\begin{equation}
T=-\frac{\Lambda r_{+}}{2 \pi}-\frac{M}{4 \pi}+\frac{Q^{2}}{6 \pi \eta r_{+}^{2}}.
\end{equation} By contrast , C is $\frac{-M }{4 \pi}$.

\section{Hawking temperature of spherically symmetric KERR-SEN black holes obtained by the RVB method}
There is a special solution (Kerr-Sen black hole solution) in a certain modified gravity, and the space-time line element of the spherically symmetric Kerr-Sen black hole can be expressed as \cite{29}
\begin{equation}
d s^{2}=-g(r) d t^{2}+\frac{d r^{2}}{g(r)}+r^{2} d \Omega^{2},
\end{equation}
where
\begin{equation}\label{Eq2}
g(r)=\left(r^{2}-2M^{\prime} r+a^{2}\right) /\left(r_{+}^{2}+a^{2}\right).
\end{equation}
 The two-dimensional line element is \cite{10,17}
\begin{equation}
d s^{2}=g(r) d \tau^{2}+\frac{d r^{2}}{g(r)},
\end{equation}
the Ricci scalar is
\begin{equation}
R=2 /\left(r_{+}^{2}+a^{2}\right).
\end{equation}
The two horizons are
\begin{equation}
r_{\pm}=M^{\prime} \pm \sqrt{M^{\prime2}-a^{2}}.
\end{equation}
The effective mass is
\begin{equation}
M^{\prime}=M-b1=M-\frac{Q^{2}}{2 M}.
\end{equation}
Note,$r_{\mp}$ is the radius of the inner and outer horizons of the black hole, and $b1$ is the parameter related to the coordinate extension.

We find the Hawking temperature in this case\cite{12,13,14}
\begin{equation}
T_{\mathrm{H}}=-\frac{1}{2}\left(\frac{1}{4 \pi} \int_{r_{-}} R d r-\frac{1}{4 \pi} \int_{r_{+}} R d r\right),
\end{equation}where $r_{-}$ is the radius of the  inner horizon, $r_{+}$ is the radius of the event horizon.
\begin{equation}
T_{H}=\frac{\sqrt{(M^{\prime})^{2}-a^{2}}}{4 \pi M^{\prime}\left(M^{\prime}+\sqrt{(M^{\prime})^{2}-a^{2}}\right)}+C,
\end{equation}C is the integration constant.
 Compared with the conventional method temperature, an important thermodynamic quantity corresponding to a black hole is the Hawking temperature $T_H$, which can be calculated at the poles as
\begin{equation}
T_{\mathrm{H}}=\left.\frac{1}{2 \pi} \lim _{r \rightarrow r_{+}} \sqrt{g^{r r}} \partial_{r} \sqrt {-g_{t t}}\right|_{\theta=0},
\end{equation}
we get 
\begin{equation}
T=\frac{\sqrt{(M^{\prime})^{2}-a^{2}}}{4 \pi M^{\prime}\left(M^{\prime}+\sqrt{(M^{\prime})^{2}-a^{2}}\right)}.
\end{equation}
By contrast, C is 0.

\section{Conclusion and Discussion}
In this work, we studied the Hawking temperature of four basic types of black holes under different f(R) gravity using the Euler characteristic in topological formulations. It was found that there is an integral constant difference between temperature results calculated by the RVB method and the method under the conventional method. In the calculation of the Hawking temperature, the integral constant can be determined from the Hawking temperature obtained by the standard definition to obtain an accurate temperature. Therefore, the topological method is also applicable to the calculation of the temperature of the black hole under the conventional f(R) gravity. It is found that the integral constants corresponding to different f(R) gravity are different. Different forms of f(R) correspond to different gravitational theories, thus their integral constants are dependent on the gravitational theory.

{\bf Acknowledgements:}\\
This work is partially supported by the National Natural Science Foundation of China(No. 11873025)


\begin{thebibliography}{99}
\bibitem[1]{1}G. W. Gibbons and R. E. Kallosh, Phys. Rev. D 51, 2839 (1995).

\bibitem[2]{2} G. W. Gibbons and S. W. Hawking, Commun. Math. Phys. 66, 291 (1979).

\bibitem[3]{3} T. Eguchi, P. B. Gilkey, and A. J. Hanson, Phys. Rep. 66, 213 (1980).

\bibitem[4]{4}  S. Liberati and G. Pollifrone, Phys. Rev. D 56, 6458 (1997).

\bibitem[5]{5}  M. Bañados, C. Teitelboim, and J. Zanelli, Phys. Rev. Lett. $\mathbf{7 2}, 957$ (1994).

\bibitem[6]{6} M. Eune, W. Kim and S. H. Yi, JHEP 03, 020 (2013).

\bibitem[7]{7}G. Gibbons, R. Kallosh and B. Kol, Phys. Rev. Lett. 77,4992(1996).

\bibitem[8]{8} D. Kubiznak and R. B. Mann, JHEP 07, 033 (2012).

\bibitem[9]{9} A. Bohr and B. R. Mottelson, ``Nuclear Structure", Vol.1 (W. A. Benjamin Inc., New York, 1969).

\bibitem[10]{10}R. K. Bhaduri, ``Models of the Nucleon", (Addison-Wesley, 1988).

\bibitem[11]{11}S. Das, P. Majumdar, R. K. Bhaduri, Class. Quant. Grav.19:2355-2368, (2002).

\bibitem[12]{12}  Robson, C.W.; Villari, L.D.M.; Biancalana, F. On the Topological Nature of the Hawking Temperature of Black Holes. Phys. Rev. D 2019, 99, 044042 .

\bibitem[13]{13}  Robson, C.W.; Villari, L.D.M.; Biancalana, F. Global Hawking Temperature of Schwarzschild-de Sitter Spacetime: a Topological Approach. arXiv:1902.02547[gr-qc].

\bibitem[14]{14}Robson, C.W.; Villari, L.D.M.; Biancalana, F. The Hawking Temperature of Anti-de Sitter Black Holes: Topology and Phase Transitions. arXiv:1903.04627[gr-qc].

\bibitem[15]{15}Zhang, Y.P.; Wei, S.W.; Liu, Y.X. Topological approach to derive the global Hawking temperature of (massive) BTZ black hole. Phys. Lett. B 2020, 810, 135788 .

\bibitem[16]{16} Xian, Junlan, and Jingyi Zhang. ``Deriving the Hawking Temperature of (Massive) Global Monopole Spacetime via a Topological Formula." Entropy 24.5 (2022): 634.

\bibitem[17]{17}Caravelli, Francesco, and Leonardo Modesto. ``Holographic effective actions from black holes." Physics Letters B 702.4 (2011): 307-311.

\bibitem[18]{18} Z. Amirabi, M. Halilsoy, S. Habib Mazharimousavi, Generation of spherically symmetric metrics in $f(R)$ gravity $[\mathrm{J}]$. The European Physical Journal C, 2016, 76(6): 338.

\bibitem[19]{19} Tan, Hongwei, et al. "The global monopole spacetime and its topological charge." Chinese Physics B 27.3 (2018): 030401.

\bibitem[20]{20}T. Multamaki, I. Vilja, Spherically symmetric solutions of modified field equations in $f(R)$ theories of gravity[J]. Physical Review D, 2006, 74(6): 064022.

\bibitem[21]{21} S.M. Carroll, V. Duvvuri, M. Trodden, M.S. Turner, Is cosmic speed-up due to new gravitational physics?[J]. Physical Review D, 2004, 70(4): 043528.

\bibitem[22]{22} S. Capozziello, V.F. Cardone, S. Carloni, A. Troisi, Curvature quintessence matched with observational data[J]. International Journal of Modern Physics D, 2003, 12(10): 1969-1982.

\bibitem[23]{23} B. Li, J.D. Barrow, The Cosmology of $f(R)$ gravity in metric variational approach[J]. Physical Review D, $2007,75(8): 084010$

\bibitem[24]{24}  L. Amendola, R. Gannouji, D. Polarski, S. Tsujikawa, Conditions for the cosmological viability of $f(R)$ dark energy models[J]. Physical Review D, 2007, 75(8): 083504.

\bibitem[25]{25}  V. Miranda, S.E. Jones, I. Waga, M. Quartin, Viable singularity-free $f(R)$ gravity without a cosmological constant[J]. Physical Review Letters, 2009, 102(22): 221101.

\bibitem[26]{26}  L. Sebastiani, S. Zerbini, Static spherically symmetric solutions in $\mathrm{F}(\mathrm{R})$ gravity[J]. The European
Physical Journal C, 2011, 71: 1591.

\bibitem[27]{27}   Younesizadeh, Younes, et al. ``What happens for the BTZ black hole solution in dilaton $f(R)$-gravity?." International Journal of Modern Physics D 30.04 (2021): 2150028.

\bibitem[28]{28} Hu, Ya-Peng, Feng Pan, and Xin-Meng Wu. ``The effects of massive graviton on the equilibrium between the black hole and radiation gas in an isolated box." Physics Letters B 772 (2017): 553-558.

\bibitem[29]{29}Ghezelbash, A. M., and H. M. Siahaan. ``Hidden and generalized conformal symmetry of Kerr–Sen spacetimes." Classical and Quantum Gravity 30.13 (2013): 135005.

\bibitem[30]{30}Hendi, S. H., B. Eslam Panah, and S. M. Mousavi. ``Some exact solutions of $f(R)$ gravity with charged (a) dS black hole interpretation." General Relativity and Gravitation 44.4 (2012): 835-853.

\bibitem[31]{31}Robson, Charles W., and Marco Ornigotti. ``Hidden Depths in a Black Hole: Surface Area Information Encoded in the ($ r $, $ t $) Sector." arXiv preprint arXiv:1911.11723 (2019).

\bibitem[32]{32}Liu, Wenbiao, and Zhao, Zheng. ``Entropy of a non-thermal equilibrium Schwarzschild-de Sitter black hole." (1995).

\bibitem[33]{33}Liu, Wenbiao, and Zhao, Zheng. ``Entropy of a non-thermal equilibrium Reissner-Nordstrom-de Sitter black hole." Journal of Beijing Normal University: Natural Sciences Edition 36.5 (2000): 626-630.

\bibitem[34]{34}Övgün, A.; Sakall,İ. Deriving Hawking Radiation via Gauss-Bonnet Theorem: An Alternative Way. Annals Phys. 2020, 413, 168071.

\bibitem[35]{35}Nashed, G. G. L. ``Rotating charged black hole spacetimes in quadratic f (R) gravitational theories." International Journal of Modern Physics D 27.07 (2018): 1850074.

\bibitem[36]{36}Zhu, Chennai, and Rong-Jia Yang. ``Horizon Thermodynamics in D-Dimensional f (R) Black Hole." Entropy 22.11 (2020): 1246.

\bibitem[37]{37}Majeed, Bushra, and Mubasher Jamil. ``Dynamics and center of mass energy of colliding particles around the black hole in f (R) gravity." International Journal of Modern Physics D 26.05 (2017): 1741017.

\bibitem[38]{38}Singha C. Thermodynamics of multi-horizon spacetimes[J]. General Relativity and Gravitation, 2022, 54(4): 1-17.

\bibitem[39]{39}Nojiri, Shin'ichi, and Sergei D. Odintsov. ``Unifying inflation with $\Lambda$CDM epoch in modified f (R) gravity consistent with Solar System tests." Physics Letters B 657.4-5 (2007): 238-245.
\end{thebibliography}
\end{document}